\documentclass[useAMS,usenatbib]{mn2e}
\usepackage{graphicx}
\title{On the masses of OJ287 black holes}
\author[M. J. Valtonen, S. Ciprini and H. J. Lehto]{M. J. Valtonen$^{1}$\thanks{E-mail:
mvaltonen2001@yahoo.com (MJV)} S. Ciprini$^{2}$ and H. J. Lehto
$^{3}$\\
$^{1}$Helsinki Institute of Physics, FIN-00014 University of Helsinki, Finland\\
$^{2}$Physics Department, University of Perugia, I-06123 Perugia, Italy\\
$^{3}$Tuorla Observatory, Department of Physics and Astronomy, University of Turku, FIN-21500 Piikki\"o, Finland}
\begin{document}

\date{Accepted . Received ; in original form}

\pagerange{\pageref{firstpage}--\pageref{lastpage}} \pubyear{2011}

\maketitle

\label{firstpage}

\begin{abstract}
Two multifrequency campaigns were carried out on OJ287 in 2005: in April when it was in its pre-outburst state, and in November, during the main 12 yr cycle outburst. The wavelength coverage was from radio to X-rays. In the optical-to-UV range the differential spectrum between the observations has a bremsstrahlung spectral shape, consistent with gas at $3 \times 10^{5}K$ temperature. Our result supports the hydrogen column density of the OJ287 host galaxy of $\sim9.3\times 10^{20} cm^{-2}$, the average value found by Gosh \& Soundararajaperumal. The $3 \times 10^{5}K$ bremsstrahlung radiation was predicted in the binary black hole model of OJ287, and it arises from a hot bubble of gas which is torn off the accretion disc by the impact of the secondary. As this radiation is not Doppler boosted, the brightness of the outburst provides an estimate for the mass of the secondary black hole, $\sim1.4\times10^{8}$ solar mass. In order to estimate the mass of the primary black hole, we ask what is the minimum mass ratio in a binary system which allows the stability of the accretion disc. By using particle simulations, we find that the ratio is $\sim1.3\times10^{2}$. This makes the minimum mass of the primary $\sim1.8\times10^{10}$ solar mass, in agreement with the mass determined from the orbit solution, $1.84 \times 10^{10}$ solar mass. With this mass value and the measured K-magnitude of the bulge of the host galaxy of OJ287, the system lies almost exactly on the previously established correlation in the black hole mass vs. K-magnitude diagramme. It supports the extension of this correlation to brighter magnitudes and to more massive black holes than has been done previously.  
\end{abstract}


\section{Introduction}
There is a fundamental relation between the central black hole mass and bulge luminosity in galaxies (Magorrian et al. 1998, Ferrarese \& Merritt 2000, G\"ultekin et al. 2009, Kormendy and Bender 2011a,b). In the study of the tightness of this correlation one should use the most accurate mass values based on the dynamics of bodies orbiting the central black hole. So far there are not many such mass determinations. The mass of the black hole in the Galactic centre gives a fixed point at the low end of the scale (Ghez et al. 2003, Genzel et al. 2010), the masses of megamaser sources fall in the middle of the range (e.g. NGC 4258, Miyoshi et al. 1995, Kuo et al. 2011), while the mass of the OJ287 primary gives the only point at the high mass end (Valtonen et al. 2010a). In addition, there are plenty of data based on the velocity dispersions of galactic centres (e.g M87, Gebhardt \& Thomas 2009), reverberation mapping measurements and finally much recent data based on emission line widths in quasars (for a review, see Ferrarese \& Ford 2005). These may be viewed as giving reliable but less accurate mass values (Valluri et al. 2004, Denney et al. 2009, 2010, Kelly et al. 2010). Even though the single data point at the high mass end is very accurate, with one percent uncertainty, one would like to have independent evidence that it is at least approximately correct. Then one could use the dynamical mass with greater confidence in correlation diagrams. Thus the primary objective in this paper is to confirm the value of the central black hole mass in OJ287.

OJ287 has a historical optical light curve extending over 120 yr from late 1800's up to date. It shows a prominent twelve year cycle (Sillanp\"a\"a et al. 1988, Valtonen et al. 2006a), with an additional two peak structure with 1 - 2 yr peak separation (Valtonen et al. 2010b). It was suggested by Sillanp\"a\"a et al. (1988) that the 12 yr structure arises from a black hole binary system where a secondary perturbs the accretion disc of the primary at regular intervals and causes indirectly increased emission in the jet of OJ287. The jet is lined up with the observer at a small viewing angle which further enforces the amplitude of brightness variations via Doppler boosting (L\"ahteenm\"aki \& Valtaoja 1999, Jorstad et al. 2005). It is extremely unlikely ($\chi^2$ probability less than $10^{-8}$) that that a single black hole model would produce such good light curve predictions as the binary model has done in the past 14 years (Valtonen et al. 2011).

This model lead to the prediction that the next major outburst after 1983 should take place in the autumn of 1994. The outburst came as scheduled (Sillanp\"a\"a et al. 1996). At this point several more detailed binary black hole models were put forward, with differing predictions for the next major outburst (see Kidger 2000). These were in two general categories: exactly periodic (non-precessing) models which predicted the main outburst in the autumn of 2006, and precessing models which gave the outburst time a year earlier (Lehto \& Valtonen 1996, Sundelius et al. 1997). There are two reasons for this. First, in the latter models the outbursts are related to the impacts of the secondary on the accretion disc of the primary. With forward precession, as in General Relativity, the impact times happen earlier than in a non-precessing periodic model. This makes about $\sim 6$ month contribution to the advance of the outburst time (Lehto \& Valtonen 1996). The second contribution comes from the fact that the accretion disc is lifted up by the approaching secondary (Sundelius et al. 1996, Ivanov et al. 1998) and the impact happens earlier than expected in a rigid disc model by another $\sim 6$ months.

In order to vertify the correctness, or otherwise, of these two clearly different models, a major monitoring campaign was set up, with frequent observations starting in early 2005. Not only was the timing important, but also the nature of the radiation at the outburst peak. Most periodic models were predicting an increase of radiation from the jet, i.e. the outburst should have a power-law like synchrotron spectrum. In contrast, the precessing binary model predicted that the radiation is bremsstrahlung from hot gas ($1 - 4\times10^{5} K$, the exact value is calculated below) which has been torn off the accretion disc. Thus, it was important to study the spectral energy distribution during the outburst, and to compare it with the spectral distribution outside the outburst time. We had two periods of XMM-Newton observations, one in April 2005 prior to the outburst, and another one in November 2005 when the outburst was expected in the precessing model. Preliminary reports of these observations have been given in Ciprini et al. (2007), Ciprini and Rizzi (2009) and Ciprini (2010).
\begin{figure}
\includegraphics[width=6cm,angle=270]{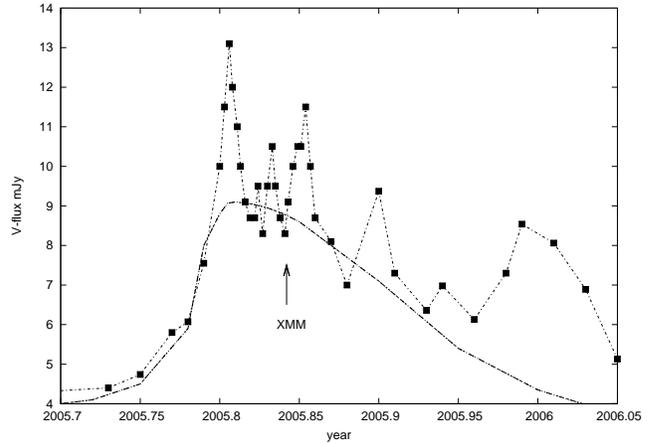} 
 \caption{The optical light curve of OJ287 during the 2005 outburst. Data points are based on Valtonen et al. (2008, 2009).}
\end{figure} 

As to the timing, the outburst followed the prediction of the precessing model (Valtonen et al. 2006b, Valtonen et al. 2008). The primary outburst came in 2005 October/November while OJ287 was in an exceptionally quiet state in the autumn of 2006 when the periodic peak brightness should have happened. Thus the result was unusually clear. However, it is also crucial to verify the second hypotheses of the precessing binary model, i.e. to see if the spectrum agrees with bremsstrahlung radiation. A major problem here would be the contribution from the ``normal'' background activity which produces secondary peaks of radiation. Luckily the second XMM-Newton period fell between two secondary peaks, and thus our measurements should tell about the nature of the primary outburst (Figure 1).

In this paper we first discuss the differential energy distribution between the two XMM-Newton observing periods, and confirm that the excess has likely bremsstrahlung nature. Since the radiation from the disc impact is not Doppler boosted, the strength of the outburst is directly linked to the mass of the secondary. We then ask what is the minimum mass of the primary which provides the stability for the accretion disc. This gives us the minimum value for the primary mass. Finally we compare this with the dynamical mass determination, and after being satisfied that the two methods give consistent results, produce a slightly modified black hole mass - K-magnitude correlation diagramme with OJ287 added to the older data.

\section{Outburst spectral energy distribution}

The two X-ray observations of OJ287 by XMM-Newton in 2005, April 12 and November 3-4 were performed during a ground-based coordinated multiwavelength (MW) campaign including both simultaneous radio/mm, near-IR and optical observations (provided mainly by the WEBT consortium\footnote{www.oato.inaf.it/blazars/webt/})
and longer term data, obtained within monitoring programs by the ENIGMA network\footnote{www.lsw.uni-heidelberg.de/projects/enigma/} facilities, and other participating observatories (see, e.g. Ciprini et al. 2007, Ciprini 2010).

In Figure 2 we plot the observations of the 2005 campaigns. The observations relating to the two instances of time are colour-coded and use different symbols. In addition, a collection of data obtained at other times are shown by grey open symbols.

For the 2005 April 12 simultaneous spectral energy distribution (SED), the spectrum from radio to X-rays followed nicely a synchrotron self-Compton (SSC) model with suitably chosen parameter values (see the discussion in Ciprini et al. 2007\footnote{the names of the 91 WEBT+ENIGMA participants in the multiwavelength campaign are found in the author list of the Ciprini et al. (2007) paper}). In contrast, the November 3-4 spectrum was quite different. The flux came up prominently in the optical/UV region, but in the radio or hard X-rays the flux remained at the pre-outburst level, making the single zone SSC model unlikely.

\begin{figure}
\includegraphics[width=\hsize,angle=0]{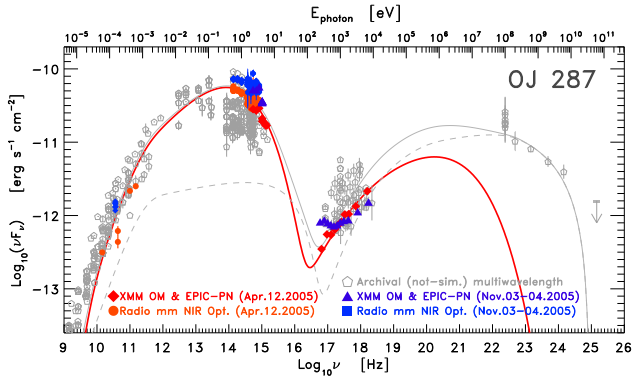}
 \caption{The SED of OJ287 during the 2005 outburst in November 3-4 (blue/dark filled triangle and square symbols), and the spectrum of the pre-burst state in 2005 April 12 (red/lighter filled diamond and circles symbols) with a pure one-zone SSC fit attempt (red/light line)). Other archival and not simultaneous data (light gray and open pentagon symbols and lines) are reported in the plot.}
\end{figure}

In Figure 3 we look at the optical/UV region in more detail, and plot the April 2005 spectrum as well as the difference spectrum (November - April) in Figure 4. The lower points (filled squares) are corrected for the Galactic absorption (Schlegel et al. 1998, Draine 2003), the upper points (crossed squares) are also corrected for the absorption in the OJ~287 system. The latter correction is important as the W1 channel, represented by the last point on the right hand side, lies right on the 2175 A absorption feature of the Galactic absortion law (Fitzpatrick 1999). Thus this point is a sensitive probe of the interstellar medium of the host galaxy of OJ~287. Other points are also affected, but to a lesser degree. The hydrogen column density of OJ~287 that brings the W1 channel point to the smooth continuation of the data from other bands is $\sim9.3\times10^{20} cm^{-2}$. Together with the hydrogen column density of $3\times10^{20} cm^{-2}$ for our Galaxy, they add up to the average of the two measured values by Gosh \& Soundararajaperumal (1995). However, we should note that this result is based on the Galactic reddening coefficient $R_V=3.1$. If this coefficient is smaller, then also the hydrogen column density needs to be smaller in order arrive at a fit similar to the one shown in Figure 4. For example, $A_V=1.1$ implies that the hydrogen column density is only half of the above mentioned value.

The April data shown in Figures 3 and 4 agree well with previous measurements from far infrared to UV . The spectrum gradually steepens from the optically thin spectrum $F_{\nu}\sim\nu^{-0.8}$ in far infrared (Impey \& Neugebauer 1988) to $F_{\nu}\sim\nu^{-1.3}$ in optical (Holmes et al. 1984, Hagen-Thorn et al. 1998, D'Arcangelo et al. 2009) to $F_{\nu}\sim\nu^{-1.7}$ in ultraviolet (Worrall et al. 1982, Maraschi et al. 1983, Gosh \& Soundararajaperumal 1995, Pursimo et al. 2000). The steepening is exactly in tune with the increasing galactic extinction from far infrared to ultraviolet in the OJ287 host galaxy, and agrees in magnitude with the extinction by the hydrogen column density around $9.3\times10^{20} cm^{-2}$. The standard extinction law (see e.g. Draine 2003, Fig. 9) for this hydrogen column density and shifted to the redshift of OJ~287 $z=0.3$ steepens the spectrum in near infrared by 0.15 units, in optical by 0.45 units and in UV between $\lambda 1100$ A and  $\lambda 3900$A by 0.9 units. Other causes for this spectral steepening are also possible (Rieke \& Kinman 1974), but the coincidence with the galactic extinction law is remarkable.

\begin{figure}
\includegraphics[width=\hsize,angle=0]{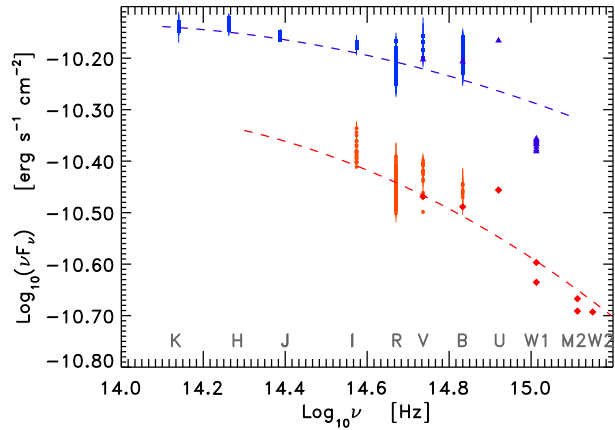}
 \caption{Near-IR, optical, near-UV simultaneous observations of OJ287 during the 2005 outburst in November 3-4 (blue/dark, top triangle symbols dashed line), compared with the observations in its pre-outburst state in 2005 April 12 (red/light bottom diamond symbols and dashed line). Data points are based on the same campaign thanks to XMM-Newton UVOT telescope observations and other ground-based (WEBT and ENIGMA observing campaigns) observatories, from K-band to W2-band.}
\end{figure}

\begin{figure}
\includegraphics[width=6cm,angle=270]{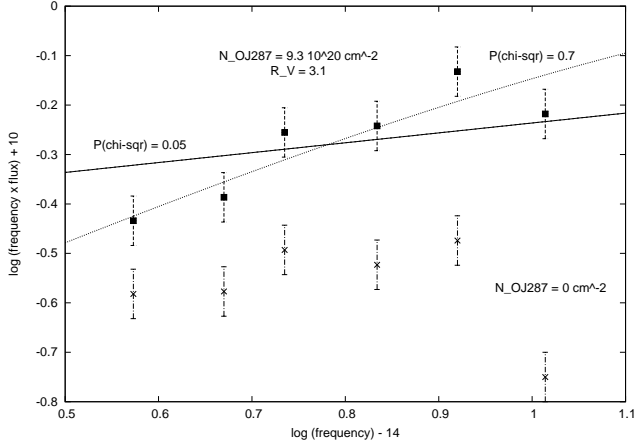}
 \caption{The difference energy spectrum of OJ287 between the outburst state and the pre-outburst state in 2005, with two-sigma errorbars. The lower points (crosses) have been corrected only for the Galactic extinction (hydrogen column density $3\times10^{20} cm^{-2}$). The upper points (filled squares) include also a correction for the extinction in the OJ287 system, corresponding to the hydrogen column density of $9.3\times10^{20} cm^{-2}$. Together they add up to the average of the two measurements by Gosh \& Soundararajaperumal (1995). A Bremsstrahlung spectral curve for $3\times10^5$ K is shown as a dotted line, with the chi-squred probability of agreement with data P = 0.7. The fit with the power-law spectrum $F_{\nu}\sim\nu^{-0.8}$ is shown by a solid line and it gives the corresponding probability of P = 0.05 which is generally considered to imply a significant deviation between the theoretical line and the data. The best power-law fit to the uncorrected points follows $F_{\nu}\sim\nu^{-1.2}$ ($\chi^2$ with a probability of agreement with data P = 0.01. The frequency unit is $10^{14}$ Hz and the unit of $\nu$$F_{\nu}$ is $10^{-10} erg cm^{-2} s^{-1}$.}
\end{figure}

We have considered two fits to the corrected points: a power-law fit with spectral index of $ \alpha=-0.4\pm0.17$ and a bremsstrahlung fit with temperature $T=3^{+6}_{-2}\times10^5K$. The power-law fit would imply a spectrum which is flatter than the normal spectrum by $\Delta\alpha\sim0.4$ units. Such flattening of the spectrum at high brightness has been previously observed in other outbursts (Zheng et al. 2008, Tavecchio et al. 2010) even though by a smaller amount. In normal flares of OJ287 the whole spectrum up to optical-UV is raised, while the inverse Compton scattering in the relativistic jet raises the X-ray flux correspondingly (Seta et al. 2009). No evidence of the inverse Compton contribution to the difference spectrum is seen in the case of the 2005 November outburst (Figure 2). 

The total luminosity of OJ287 from radio to UV would be about $2\times10^{46}$ erg/s if its radiation is isotropic (Worrall et al. 1982). However, the radiation is strongly beamed to our direction (Sitko and Junkkarinen 1985, Jorstad et al. 2005), and thus the total luminosity, integrated over all directions, may be only of the order $2\times10^{45}$ erg/s. It implies a rather low accretion rate of $\sim 10^{-3}$ of the Eddington rate $\dot{M}_{Edd}$ for the primary mass of $M=1.84\times10^{10}$ solar mass.

 In the binary model the secondary impacts the accretion disc at the distance of 0.065 pc from the primary in 2005 (Valtonen 2007). Ivanov et al. (1998) estimate that the peak bremsstrahlung luminosity arising from the impact is about equal to the Eddington luminosity of the impacting black hole, i.e $\sim 2\times10^{46}$ erg/s for a $M_{sec}=1.4\times10^{8}$ solar mass secondary, equal to the total (beamed) luminosity of OJ287. This radiation does not arise from accretion of matter to the secondary black hole but rather from a bubble of hot gas which is not accreted by it. In Section 5 we will return to the question of how much radiation may arise from the accretion process to the $M_{sec}=1.4\times10^{8}$ solar mass secondary. A more detailed calculation was carried out by Lehto \& Valtonen (1996). Scaling the accretion rate to $10^{-3}\dot{M}_{Edd}$ where $\dot{M}_{Edd}$ is the critical accretion rate of the primary black hole, and using the scaling laws of Stella and Rosner (1984) for the $\alpha_g$ disk model of Sakimoto and Coroniti (1981), we find that the expected peak luminosity is $10^{46}\dot{M}^{0.45}$ erg/s, i.e. roughly one half of the observed total luminosity.

This extra luminosity is concentrated rather strongly in optical-UV. The peak bremsstrahlung flux at the optical V-band becomes $S\sim4 \alpha_g^{-0.8}\dot{M}^{0.12}M_{sec}$ mJy where $M_{sec}$ is the mass of the impacting (secondary) black hole in units of $1.4\times10^8$ solar mass. For a typical $\alpha_g$ value from observations (Dubus, Hamery and Lasota 2004, Qiao and Liu 2009) of $\alpha_g\sim0.3$, $S\sim10M_{sec}$mJy. This may be compared with the observed $S\sim6$ mJy which agrees well with the calculated value when we take account of the likely extinction (Figure 4), and use our reference value for the secondary black hole mass of $1.4\times10^8$ solar mass. 

The disc at the impact point is heated initially to $\sim5\times10^6$K. The bubble torn off from the disc has the initial optical depth $\tau\sim60\alpha_g^{-0.85}\dot{M}^{0.3}M_{sec}^{1.44}$, using the previous scaling for $\dot{M}$ and $M_{sec}$. The expansion of the bubble by a factor of $\tau^{4/7}\sim10 \alpha_g^{-0.49}\dot{M}^{0.17}M_{sec}^{0.82}$ is required in order to make $\tau=1$ at the time of the outburst. In adiabatic expansion the radiation dominated gas becomes cooler, and acquires the temperature of $T\sim 5\times10^{5} \alpha_g^{0.49}\dot{M}^{-0.17}M_{sec}^{-0.82}$K. Our fit to the energy spectrum in Figure 4 is fully consistent with this interpretation with $\alpha_g\sim0.3$, $\dot{M}\sim10^{-3}\dot{M}_{Edd}$ and $M_{sec}\sim1.4\times10^{8}$ solar mass.

Thus the mass value determined from the orbit solution $M_{sec}\sim1.4\times10^8$ solar mass agrees well with the measured properties of the optical outburst.  Our next step is to ask how much bigger the primary mass has to be in order that it can guarantee the stability of the primary accretion disc.

\section{Stability of the disc}
The stability of accretion discs has been studied by many authors, in particular for the purposes of modelling interacting binary stars (e.g. Fragner \& Nelson 2010). Larwood et al. (1996) studied different inclinations and binary mass ratios while Liu \& Shapiro (2010) consider a wide range of mass ratios, but only coplanar orbits. The model for OJ~287 has a binary orbit which is close to perpendicular to the disc plane and which has a large mass ratio. The stability of such systems has not been studied previously.

We have modeled the accretion disc by a system of 6800 particles which gives an adequate resolution in the study of the loss of disc particles due to the binary action. The particles are placed initially in a single plane approximately perpendicular to the binary orbit. The orbit integration is started around year 1850 and continued until year 2050 using the latest OJ287 binary model parameters which include the effects of the binary spin (Valtonen 2010a). The orbital elements of the particles are calculated at each time step. In the astrophysical disc model of Lehto \& Valtonen (1996), the disc thickness is such that orbits with inclinations less than $6^{\circ}$ remain in the disc (see Figure 4 of Valtonen 2007 for the standard disc profile).

We varied the secondary mass $m_2$ in order to get a set of runs with different mass ratios of binary black hole components, and for each run we calculate the fraction $f$ of particles lost from the disc, i.e. particles with orbital inclinations greater that $6^{\circ}$. Then the lifetime of the disc is determined from $t_{life}$=$200 yr/f$. The results are illustrated in Figure 5.

The disc lifetime should be compared with the disc replenishment time which occurs in the viscous time scale (Ivanov et al. 1999):

\begin{equation}
                t_{visc}\sim10^{7}\alpha^{-0.8}\dot{M}^{-0.4}r^{1.4} yr
\end{equation}

where $\alpha$ is the viscosity parameter, $\dot{M}$ is the accretion rate in units of $0.01\dot{M}_{Edd}$, $\dot{M}_{Edd}$ is the rate which gives the Eddington luminosity for the $M=1.84\times10^{10}$ solar mass central black hole, and $r$ is the outer radius of the disc which has been perturbed, in units of 20 Schwarzschild radii. The parameters of OJ287 are such that $t_{visc}\sim10^{7}$ yr.

We notice that the disc particles are lost quite fast unless the mass ratio $M_r$ is at least one hundred. The typical replenishment time scale of the disc by the viscous accretion process, $t_{life}\sim10^7$ yr, corresponds to the stability ratio $\sim130$. This makes the minimum value of the primary mass $\sim1.8\times10^{10}$ solar mass which is quite close to the mass determined from the orbit, $1.84\times10^{10}$ solar mass. Liu \& Shapiro (2010) find that the instability gap develops in the disk when the mass ratio $M_{r}=5\times10^{2}$ or less. The difference is in the expected direction since tidal effects work less efficiently at high inclination than in the coplanar case (Ivanov et al. 1999).
\begin{figure}
\includegraphics[width=6cm,angle=270]{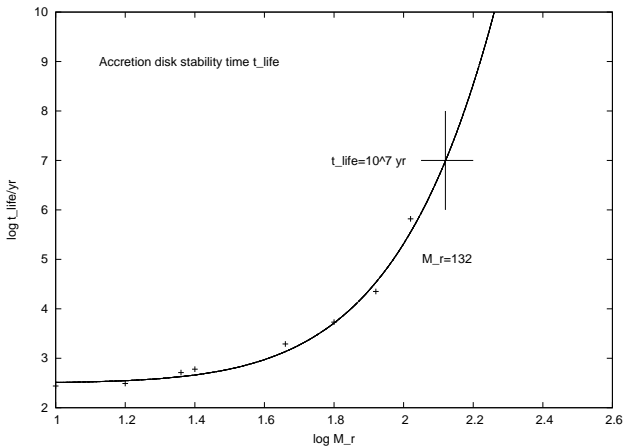}
 \caption{The stability time $t_{life}$ of an accretion disc when the mass ratio $M_r$ is varied.}
\end{figure}

\section{Black hole mass - Bulge luminosity correlation}
 Wright et al. (1998) measured the absolute magnitude of the bulge of the OJ287 host galaxy as $M_K=-28.9\pm0.6$. In Figure 6 we plot OJ287 in the black hole mass -  K-magnitude diagramme using this value of $M_{K}$ and the dynamical mass from Valtonen et al. (2010b). The other points are from a recent sample of Kormendy \& Bender (2011a). The line of regression as well as the one sigma parallels are also drawn. We note that the point representing OJ287, with the dynamical mass from Valtonen et al. (2010b) and the K-magnitude from Wright et al. (1998) falls exactly on the regression line.

Wright et al. (1998) derived also a lower limit for the host galaxy magnitude in the R-band and found that it is consistent with the R-band measurements by Hutchings (1987), Heidt et al. (1999) and Pursimo et al. (2002). Wright et al. (1998) point out that the K-band and R-band magnitudes imply about two magnitudes of extinction in the OJ287 host galaxy in the R-band, and larger hydrogen column density that what we have discussed above. However, there does not need to be an exact correspondense between the measurements from the quasar and from the outer parts of the host galaxy since the lines of sight are different. Also there are large variations in extinction laws between different galaxies (Falco et al. 1999). It is quite likely that OJ287 is a recent merger considering its double black hole nature (Iwasawa et al. 2010), and this impression is given also by the perturbed image of the host galaxy (Benitez et al. 1996). Thus about two magnitudes of extinction in the R-band is not out of question since mergers tend to be dusty (Sanders et al. 1988). New observations of the OJ287 host galaxy in different colours would clarify the situation.

\begin{figure}
\includegraphics[width=6cm,angle=270]{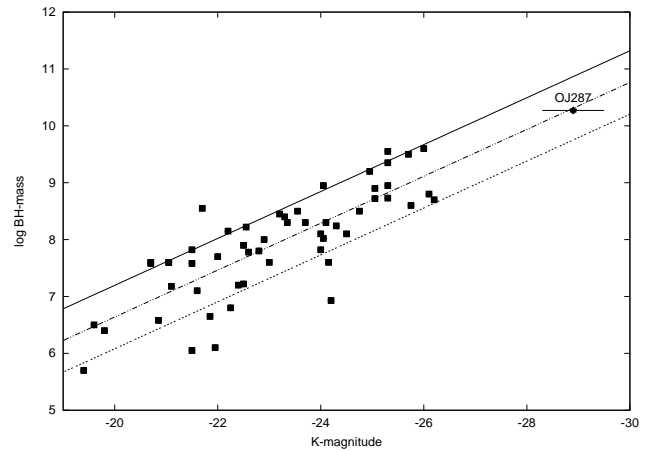} 
 \caption{The black hole mass - K-magnitude correlation in the sample of Kormendy \& Bender (2011), squares, and the OJ287 point with magnitude error bars.
 The lines are the best fit regression to the Kormendy \& Bender sample and the two sigma deviations from it.}
\end{figure} 
\section{Variability studies}
Studies of short time scale variability have been used to set limits on the black hole masses in OJ287 and in other quasars (Gupta et al. 2012). The idea is that the shortest periodic variability component may be related to the period of the innermost stable orbit of the radiating gas falling into the black hole. The same time scale, but shortened by the relativistic Doppler boosting factor, may also appear in the jet emission (Sagar et al. 2004).

In case of OJ287 we have two time scales to consider: the scale related to the primary and the scale related to the secondary. Why should we consider the light output from the secondary whose mass is expected to be two orders of magnitude smaller than the mass of the primary? So far we have not resolved the two components separately but only record the total output from the two sources.
Sundelius et al. (1997) calculated the infall rates of matter to both black holes, and found that the peak rates were about the same. The expected flux depends also on other factors such as the mass and the spin of the black hole. If the spin of the secondary is close to maximal, as would be natural after repeated accretion events from the constant high angular momentum source of gas (the primary disc), its luminosity may be an order of magnitude greater than the corresponding luminosity for a slowly rotating black hole. And since the primary black hole is likely to spin rather slowly (Valtonen et al. 2010a), the flux from the secondary could easily amount to about $10\%$ of the flux of the primary. As to the Doppler boosting, it could be significant also in the secondary, since its spin axis is likely to be parallel to the angular momentum vector of its source of accreting gas, the primary disk. Thus the jet of the secondary could be beamed towards us just as the jet of the primary is (Valtonen and Wiik 2012).

To find evidence for the emission of the secondary, we need to look at short time scale variability in OJ287. It has been found to be variable from 15 minutes time scale upwards on many occasions (Valtaoja et al. 1985, Dultzin-Hacyan et al. 1997, Gupta et al. 2012), and on one occasion the light curve has shown sinusoidal variations of the period of 228 minutes (Sagar et al. 2004). If this variation is associated with the last stable orbit of a maximally rotating black hole, the mass of the black hole is $1.46\times10^8$ solar mass (Gupta et al. 2012), i.e. identical to the mass obtained from the orbit solution (Valtonen et al. 2010b). On the other hand, when the same frequency of variations is transmitted to the jet and is observed, the associated time scale is shorter by a factor of about 18.7, the Lorentz factor of the jet in the optical region. This is consistent with shortest observed variability time scale of 15.7 minutes in OJ287.

Neronov and Vovk (2011) associate the 2009 October-November gamma-ray flare of OJ287 also with the secondary black hole.

\begin{figure}
\includegraphics[width=6cm,angle=270]{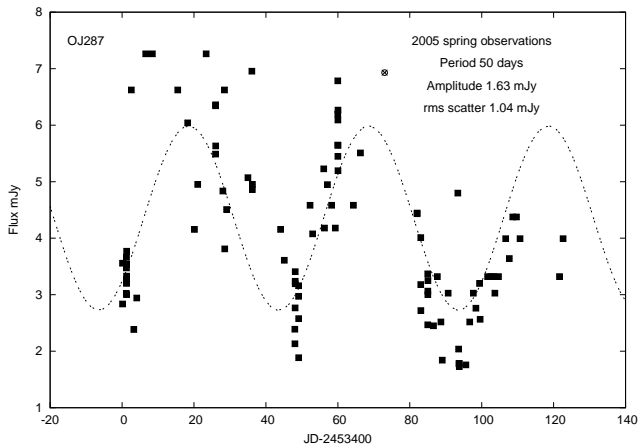}
 \caption{Observations of OJ287 brightness in the spring of 2005 (squares), including the pre-outburst point in April 12 (crossed circle), compared with a 50 day periodic cyclic pattern.}
\end{figure}

As to the primary mass, the estimated period of the last stable orbit is around 100 days, the exact value depending on the spin of the black hole. The perturbation of matter from the last stable orbit, leading to its transfer to the jet, is affected by the tidal force of the companion.  Since the tidal force is symmetric relative to the central black hole, the tidal transfer period from the last stable orbit should be one half of this value, i.e. about 50 days. Wu et al. (2006) reported the periodicity of 46 days in OJ287 during one observing season in the spring of 2005. The analysis of the 14 year stretch of optical monitoring data from 1996 to 2010 supports this finding (Valtonen et al. 2011).

 We have added 41 points from the 2005-2010 optical monitoring campaign of OJ287 (Valtonen et al. 2009) to the 61 points of Wu et al. (2006) for the spring 2005; they are plotted in Figure 7. There is a clear variation at about 50 day period in this segment. A similar periodic variation extends up to January 2008; the peaks at 2005.85 and 2006.0 are the high points of this cyclic variation (Figure 1). The April 2006 outburst and the three highest outburst peaks in late 2007 and early 2008 (excluding the 2007 September impact outburst) are also part of the same 50 day cycle (see Figure 3 of Valtonen et al. 2009). Within statistical uncertainties, it is possible that the period of one half of the last stable orbit has been detected, and that the primary mass is confirmed also in this way.
            
\section{Discussion}
We find that the mass values for the two black holes in the OJ287 system, as derived from the orbital solution, are quite consistent with values obtained from astrophysical arguments. The chain of arguments starts from identifying the 2005 outburst radiation as bremsstrahlung. Since its likely origin is the collision of the secondary on the accretion disc of the primary, this allows the determination of the mass of the secondary. The requirement that the disc is stable inspite of the binary action puts a lower limit on the mass of the primary. It is very near the mass determined from the orbit solution. Thus we can have increased confidence in the correctness of the orbit solution.

We place OJ287 in the  black hole mass vs. K-magnitude  diagramme and find that it lies almost exactly on the mean line determined from the lower black hole masses. This gives us confidence that this relation can be extended to up to the largest measured black hole masses.

As a side result we find that the host galaxy of OJ287 has a moderate amount of extinction which corresponds to the hydrogen column density of $\sim9.3\times10^{20} cm^{-2}$, similar to values measured in some other galaxies from the 2175 A extinction feature (Wang et al. 2004). The dust and the perturbed appearance of the host both suggest that OJ287 has acquired its second black hole in a merger of two galaxies, possibly leading to binary black hole evolution as calculated by Iwasawa et al. (2010).

The observed periodicities in the light curves of OJ287 are consistent with a model where both the slowly spinning primary black hole and the fast spinning secondary black hole contribute to the optical light variations (Neronov and Vovk 2011). In future it may become possible to see the orbital motion directly when the required microarcsecond resolution is achieved. It may also be possible to detect the motion of the secondary from its spectral lines. The orbital speed of the secondary is so fast that its spectral lines are shifted very far from their normal positions, outside the spectral region where the line searches have been carried out (Nilsson et al. 2010).
 
\section*{Acknowledgments}
This work is based on observations obtained with XMM-Newton, an ESA science mission with instruments and contributions directly funded by ESA member states and NASA. We also acknowledge funding from European Community's Human Potential Programme under contract HPRN-CT-2002-00321. This work is partly based on data taken and assembled by the WEBT Collaboration and stored in the WEBT archive at the INAF Observatory of Torino. This work is also partly based on an ENIGMA-related observing campaign. ENIGMA is the European Network for the Investigation of Galactic Nuclei through Multifrequency Analysis, representing a network on Blazar research, funded by the European Commission through the Training and Mobility through Research programme.

\label{lastpage}


\begin{thebibliography}{99}
\bibitem[\protect\citeauthoryear{Benitez et al.}{1996}]{b1} Benitez, E., Dultzin-Hacyan, D., Heidt, J., Sillanp\"a\"a, A., Nilsson, K., Pursimo, T., Teerikorpi, P. \& Takalo, L. O., 1996, ApJ, 464, L47
\bibitem[\protect\citeauthoryear{Ciprini et al.}{2007}]{c1} Ciprini, S. et al., 2007,
MemSAI, 78, 781
\bibitem[\protect\citeauthoryear{Ciprini \& Rizzi, N.}{2008}]{c2} Ciprini, S. \& Rizzi, N., 2008,
in Proceedings of the Workshop on Blazar Variability across Electromagnetic Spectrum, April 22-25, 2008, Palaiseau, France, p. 30
\bibitem[\protect\citeauthoryear{Ciprini}{2010}]{c3} Ciprini, S., 2010,
in X-ray astronomy 2009, present status, multi-wavelength approach and future perspectives, AIP Conf.Proc., 1248, p. 411
\bibitem[\protect\citeauthoryear{D'Arcangelo et al.}{2009}]{d1} D'Arcangelo, F. D. et al., 2019, ApJ, 697, 985
\bibitem[\protect\citeauthoryear{Dultzin-Hacyan et al.}{1997}]{d2} Dultzin-Hacyan, D., Takalo, L. O., Benitez, E., Sillanp\"a\"a, A., Pursimo, T., Lehto, H. \& de Diego, J. A., 1977,	RevMexA\&A, 33, 17	
\bibitem[\protect\citeauthoryear{Denney et al.}{2009}]{d3} Denney, K. D., Peterson, B. M., Dietrich, M., Vestergaard, M. \& Bentz, M. C., 2009, ApJ, 692, 246
\bibitem[\protect\citeauthoryear{Denney et al.}{2010}]{d4} Denney, K. D. et al., 2019, ApJ, 721, 715
\bibitem[\protect\citeauthoryear{Draine}{2003}]{d5} Draine, B. T., 2003, ARA\&A, 41, 241
\bibitem[\protect\citeauthoryear{Dubus et al.}{2004}]{d6} Dubus, G., Hamery, J.-M. \& Lasota, J.-P., 2004, A\&A, 373, 251
\bibitem[\protect\citeauthoryear{Falco et al.}{1999}]{f1} Falco, E. E. et al., 1999, ApJ, 523, 617
\bibitem[\protect\citeauthoryear{Ferrarese\& Ford}{2005}]{f2} Ferrarese, L. \& Ford, H., 2005,
SpaceSciRev, 116, 523
\bibitem[\protect\citeauthoryear{Ferrarese\& Merritt}{2000}]{f3} Ferrarese, L. \& Merritt, D., 2000,
ApJ, 539, L9
\bibitem[\protect\citeauthoryear{Fitzpatrick}{1999}]{f4} Fitzpatrick, E. L. , 1999, PASJ, 111, 63
\bibitem[\protect\citeauthoryear{Fragner\& Nelson}{2010}]{f5} Fragner, M. M. \& Nelson, R. P., 2010,
A\&A, 511, 77
\bibitem[\protect\citeauthoryear{Gebhardt \& Thomas}{2009}]{g1} Gebhardt, K. \& Thomas, J., 2009,
ApJ, 700, 1690
\bibitem[\protect\citeauthoryear{Genzel et al.}{2010}]{g2} Genzel, R., Eisenhauer, F. \& Gillessen, S., 2010, RevModPhys, 82, 3121
\bibitem[\protect\citeauthoryear{Ghosh\& Soundararajaperumal}{1995}]{g3} Ghosh, K. K. \& Soundararajaperumal, S., 1995,
ApJS, 100, 37
\bibitem[\protect\citeauthoryear{G\"ultekin et al.}{2009}]{g4} G\"ultekin et al., 2009,
ApJ, 698, 198
\bibitem[\protect\citeauthoryear{Gupta et al.}{2012}]{g5} Gupta, S. P., Pandey, U. S., Singh, K., Rani, B., Pan, J., Fan, J. H. \& Gupta, A. C., 2012, New Astronomy, 17, 8
\bibitem[\protect\citeauthoryear{Hagen-Thorn et al.}{1998}]{h1} Hagen-Thorn, V. A., Marchenko, S. G., Takalo, L. O., Sillanp\"a\"a, A., Pursimo, T., Boltwood, P., Kidger, M. \& Gonzalez-Perez, J. N., 1998,
A\&AS, 133, 353
\bibitem[\protect\citeauthoryear{Heidt et al.}{2007}]{h2} Heidt, J. et al., 1999,
A\&A, 352, L11
\bibitem[\protect\citeauthoryear{Holmes et al.}{1984}]{h3} Holmes, P.A. et al., 1984,
MNRAS, 211, 497
\bibitem[\protect\citeauthoryear{Hutchings}{1987}]{h4} Hutchings, J. B. , 1987,
ApJ, 320, 122
\bibitem[\protect\citeauthoryear{Ivanov et al.}{1998}]{i1} Ivanov, P. B., Igumenshchev, I. V. \& Novikov, I. D., 1998, ApJ, 507, 131
\bibitem[\protect\citeauthoryear{Impey \& Neugebauer}{1988}]{i2} Impey, C. D. \& Neugebauer, G., 1988, AJ, 95, 307
\bibitem[\protect\citeauthoryear{Ivanov et al.}{1999}]{i3} Ivanov, P. B., Papaloizou, J. C. B. \& Polnarev, A. G., 1999, MNRAS, 307, 79
\bibitem[\protect\citeauthoryear{Iwasawa et al.}{2010}]{i4} Iwasawa, M., An, S., Matsubayashi, T., Funato, Y. \& Makino, J., 2010,
arXiv:1011.4017
\bibitem[\protect\citeauthoryear{Jorstad et al.}{2005}]{j1} Jorstad, S. G. et al., 2005, AJ, 130, 1418
\bibitem[\protect\citeauthoryear{Kelly et al.}{2010}]{k1} Kelly, B. C., Vestergaard, M., Fan, X., Hopkins, P., Hernquist, L. \& Siemiginowska, A., 2010, ApJ, 719, 1315
\bibitem[\protect\citeauthoryear{Kidger}{2000}]{k2} Kidger, M., 2000, AJ, 119, 2053
\bibitem[\protect\citeauthoryear{Kormendy \& Bender}{2011}]{k3} Kormendy, J. \& Bender, R., 2011a,
Nature, 469, 374 
\bibitem[\protect\citeauthoryear{Kormendy \& Bender}{2011}]{k4} Kormendy, J. \& Bender, R., 2011b,
Nature, 469, 377 
\bibitem[\protect\citeauthoryear{Kuo et al.}{2011}]{k5} Kuo, C. Y. et al., 2011, ApJ, 727, 20
\bibitem[\protect\citeauthoryear{L\"ahteenm\"aki \& Valtaoja}{1999}]{l1} L\"ahteenm\"aki \& Valtaoja, E., 1999, ApJ, 521, 493
\bibitem[\protect\citeauthoryear{Larwood et al.}{1996}]{l2} Larwood, J. D., Nelson, R. P.,  Papaloizou, J. C. B. \& Terquem, 1996, MNRAS, 282, 597
\bibitem[\protect\citeauthoryear{Lehto \& Valtonen}{1996}]{l3} Lehto, H. J. \& Valtonen, M. J., 1996,
ApJ, 460, 207
\bibitem[\protect\citeauthoryear{Liu \& Shapiro}{2010}]{l4} Liu, Y. T. \& Shapiro, S. L., 2010, PhRvD, 82, 3011
\bibitem[\protect\citeauthoryear{Magorrian et al.}{1998}]{m1} Magorrian, J. et al., 1998, AJ, 115, 2285
\bibitem[\protect\citeauthoryear{Maraschi et al.}{1983}]{m2} Maraschi, L., Tanzi, E. G., Treves, A. \& Falomo, R., 1983, A\&A, 127, L17
\bibitem[\protect\citeauthoryear{Miyoshi et al.}{1995}]{m3} Miyoshi, M., Moran, J., Herrnstein, J., Greenhill, L., Nakai, N., Diamond, P. \& Inoue, M., 1995, Nature, 373, 127
\bibitem[\protect\citeauthoryear{Neronov \& Vovk}{2011}]{n1} Neronov, A. \& Vovk, Ie., 2011, MNRAS, 412, 1389
\bibitem[\protect\citeauthoryear{Nilsson et al.}{2010}]{n2} Nilsson, K., Takalo, L. O., Lehto, H. J.\& Sillanp\"a\"a, A., 2010, A\&A, 516, 60	
\bibitem[\protect\citeauthoryear{Pursimo et al.}{2000}]{p1} Pursimo, T. et al., 2000,
A\&AS, 146, 141
\bibitem[\protect\citeauthoryear{Pursimo et al.}{2002}]{p2} Pursimo, T., Nilsson, K., Takalo, L. O., Sillanp\"a\"a, A., Heidt, J. \& Pietil\"a, H., 2002,
A\&A, 381, 810
\bibitem[\protect\citeauthoryear{Qiao \& Liu}{2009}]{q1} Qiao, E. \& Liu, B.F., 2009, PASJ, 61, 403
\bibitem[\protect\citeauthoryear{Rieke \& Kinman}{1974}]{r1} Rieke, G. H. \& Kinman, T. D., 1974, ApJ, 192, L115
\bibitem[\protect\citeauthoryear{Sagar et al.}{2004}]{s1} Sagar, R., Stalin, C. S., Gopal-Krishna, Sanders \& Wiita, P. J., 2004, MNRAS, 348, 176
\bibitem[\protect\citeauthoryear{Sakimoto \& Coroniti}{1981}]{s2} Sakimoto, P. J. \& Coroniti, F. V., 1981, ApJ, 247, 19
\bibitem[\protect\citeauthoryear{Sanders et al.}{1988}]{s3} Sanders, D. B., Soifer, B. T., Elias, J. H., Neugebauer, G. \& Matthews, K., 1988, ApJ, 328, 35	
\bibitem[\protect\citeauthoryear{Schlegel et al.}{1998}]{s4} Schlegel, D. J., Finkbeiner, D. P. \& Davis, M., 1998, ApJ, 500, 525	
\bibitem[\protect\citeauthoryear{Seta et al.}{2009}]{s5} Seta, H. et al., 2009, PASJ, 61, 1011 
\bibitem[\protect\citeauthoryear{Sillanp\"a\"a et al.}{1988}]{s6} Sillanp\"a\"a, Haarala, S., Valtonen, M. J., Sundelius, B. \& Byrd, G. G., 1988, ApJ,
325, 628
\bibitem[\protect\citeauthoryear{Sillanp\"a\"a et al.}{1996}]{s7} Sillanp\"a\"a et al., 1996, A\&A,
305, L17
\bibitem[\protect\citeauthoryear{Stella \& Rosner}{1984}]{s8} Stella, L. \& Rosner, R., 1984,
ApJ, 277, 312
\bibitem[\protect\citeauthoryear{Sundelius et al.}{1996}]{s9} Sundelius, B., Wahde, M., Lehto, H. J. \& Valtonen, M. J., 1996, in Blazar Continuum Variability, ASP Conf. Ser., 110, 99
\bibitem[\protect\citeauthoryear{Sundelius et al.}{1997}]{s10} Sundelius, B., Wahde, M., Lehto, H. J. \& Valtonen, M. J., 1997, ApJ,
484, 180
\bibitem[\protect\citeauthoryear{Tavecchio et al.}{2010}]{t1} Tavecchio, F., Ghisellini, G., Ghirlanda, G., Foschini, L. \& Maraschi, L., 2010, MNRAS,
401, 1570
\bibitem[\protect\citeauthoryear{Valluri et al.}{2004}]{v1} Valluri, M., Merritt, D. \& Emsellem, E., 2004, ApJ,
602, 66
\bibitem[\protect\citeauthoryear{Valtonen et al.}{2006}]{v2} Valtonen, M. J. et al., 2006, ApJ, 646, 36
\bibitem[\protect\citeauthoryear{Valtonen}{2007}]{v3} Valtonen, 2007, ApJ, 659, 1074
\bibitem[\protect\citeauthoryear{Valtonen et al.}{2008}]{v4} Valtonen, M. J., Kidger, M., Lehto, H. J. \& Poyner, G., 2008, A\&A, 477, 407
\bibitem[\protect\citeauthoryear{Valtonen et al.}{2009}]{v5} Valtonen, M. J. et al., 2009, ApJ, 698, 781
\bibitem[\protect\citeauthoryear{Valtonen et al.}{2010a}]{v6} Valtonen, M. J. et al., 2010a, ApJ, 709, 725

\bibitem[\protect\citeauthoryear{Valtonen et al.}{2010b}]{v7} Valtonen, M. J. et al., 2010b, CeMDA, 106, 235
\bibitem[\protect\citeauthoryear{Valtonen et al.}{2011}]{v8} Valtonen, M. J., Lehto, H. J., Takalo, L. O. \& Sillanp\"a\"a, A., 2011,
ApJ, 729, 33 
\bibitem[\protect\citeauthoryear{Valtonen \& Wiik}{2012}]{v9} Valtonen, M. J. \& Wiik, K., 2012, MNRAS, in press
\bibitem[\protect\citeauthoryear{Wright et al.}{1998}]{w3} Wright, S. C., McHardy, I. M. \& Abraham, R. G., 1998, MNRAS,
295, 799
\bibitem[\protect\citeauthoryear{Zheng et al.}{2008}]{z1} Zheng, Y.G., Zhang, X., Bi, X.W., Hao, J.M. \& Zhang, H.J., 2008, MNRAS,
385, 823
\end{thebibliography}
\end{document}